\documentclass{eigenorbit}

\newcommand{\ek}{\texttt{eigenkit}}

\title{eigenkit: A Provider-Neutral Quantum Computing Library}
\shorttitle{eigenkit: a provider-neutral quantum computing library}

\addauthor{Laia Domingo}{1}
\addauthor{Daniil Tarpanov}{1}
\addauthor{Arsenii Senokosov}{1}
\addauthor{Asel Sagingalieva}{1}

\addaffiliation{1}{EigenOrbit S\`arl, Luxembourg}

\begin{document}

\begin{abstract}
Quantum software ecosystems provide increasingly capable tools for simulation and hardware execution, but differences in programming models, supported features, and backend capabilities can make it difficult to move and systematically compare the same quantum algorithm across execution environments. We introduce \ek, a provider-neutral quantum computing library that separates algorithm definition from backend execution through a common circuit representation and provider interface. The library currently supports 12 execution providers spanning simulators and quantum hardware, provides backend-independent gradient computation, and integrates with standard scientific-computing and machine-learning workflows through NumPy, PyTorch, and scikit-learn interfaces. We evaluate the practical relevance of this abstraction across both simulation and quantum hardware. Simulator performance varies substantially with circuit size, structure, batching, and application workload, showing that no single backend is consistently preferable. On quantum hardware, experiments on IBM Heron and IQM Emerald show that which processor is better suited to a given algorithm depends not only on hardware constraints, but also on how well the topology of the quantum algorithm matches the physical connectivity of the processor. These results demonstrate that backend portability is not only a software-design convenience, but can be important for both computational efficiency and quantum execution quality.
\end{abstract}

\maketitle

%% =====================================================================
\section{Introduction}
\label{sec:intro}

As quantum computing moves from isolated algorithm demonstrations toward practical experimentation, quantum software becomes the layer that connects algorithm design, classical benchmarking, and execution on simulation and heterogeneous quantum hardware. Once a quantum algorithm has been designed, two practical questions immediately arise: where should it be executed, and how should its performance be evaluated? For execution, users can choose among a growing number of simulators and quantum processing units (QPUs). Simulators differ in supported functionality, scalability, and performance across workloads, while QPUs offer different trade-offs in system size, connectivity, native operations, coherence, and gate fidelities. There is therefore no universally preferable backend, and comparing alternatives is often necessary. This diversity is substantial enough that heterogeneous quantum architectures are themselves being explored to combine complementary hardware capabilities~\cite{darpa_harq}. At the same time, evaluating a quantum algorithm typically requires integrating it with, or benchmarking it against, classical models implemented in established scientific and machine-learning libraries such as NumPy, PyTorch, and scikit-learn~\cite{numpy,pytorch,sklearn}.

A growing ecosystem of quantum software frameworks addresses much of this workflow, including Qiskit, PennyLane, CUDA-Q, TensorCircuit, Qulacs, qsim, and Amazon Braket~\cite{qiskit,pennylane,cudaq,tensorcircuit,qulacs,qsim,braket}. These frameworks share many common goals and provide substantial overlap in capabilities, but differ in their programming models, backend support, and the extent to which specific features are available. For example, some place greater emphasis on high-performance simulation, others on hardware access or hybrid quantum-classical workflows, and support for functionalities such as gradient computation is not uniform across all backends. As a result, moving the same algorithm between execution environments can still require changes to circuit definitions, provider-specific code, or differentiation strategies, while integration with classical machine-learning workflows is handled differently across frameworks. This makes systematic comparison across backends cumbersome, precisely when the choice of execution environment can have a substantial impact on performance, as we show below for both simulation and quantum hardware.

We introduce \ek, a provider-neutral quantum computing library designed to separate the definition of a quantum algorithm from the backend used to execute it. The library provides a common interface across heterogeneous simulation and hardware environments while remaining compatible with standard scientific-computing and machine-learning workflows. Its goal is not to replace existing quantum frameworks, but to provide an abstraction layer that makes algorithms easier to move, compare, differentiate, and integrate across them. The main contributions of \ek{} are:
\begin{itemize}
  \item \textbf{Provider-neutral execution.} A circuit can be defined once and executed, with minimal changes to user code, across multiple simulation backends, including an in-house numpy simulator, PennyLane, Qiskit Aer, CUDA-Q, TensorCircuit, Qulacs, and qsim, as well as QPUs accessed through Qiskit Runtime, IQM Resonance, Quandela's Perceval and Amazon Braket.
  \item \textbf{Backend-independent differentiation.} The library exposes a common gradient interface across execution backends, using native differentiation when available and library-level parameter-shift methods otherwise. This allows differentiable quantum models to be moved between backends without changing their optimization workflow.
  \item \textbf{Integration with the classical scientific stack.} Quantum models can interface with NumPy and PyTorch, while higher-level estimators can follow scikit-learn-style \texttt{fit}/\texttt{predict} workflows, allowing quantum components to be incorporated directly into classical machine-learning pipelines.
  \item \textbf{High-level algorithmic functionality.} The library includes implementations and utilities for several application domains, including quantum neural networks, quantum reservoir computing, and quantum support vector machines for quantum machine learning; quantum optimization algorithms; quantum chemistry utilities and optimization; and algorithms such as iterative quantum amplitude estimation and Harrow--Hassidim--Lloyd (HHL), see \cref{sec:code}.
\end{itemize}

The remainder of this paper is organized as follows. \Cref{sec:design} describes the design and methodology of \ek, including its lightweight and modular dependency structure, common internal representation, provider interface, compilation and dispatch mechanisms, extensibility, and backend-independent differentiation. \Cref{sec:experiments} evaluates the practical implications of this abstraction through simulation and QPU experiments, showing that the most suitable backend depends strongly on the workload and that provider switching can therefore be important in practice. \Cref{sec:limitations} discusses the current limitations of the library, and \cref{sec:conclusions} concludes the paper.

%% =====================================================================
\section{Design}
\label{sec:design}

This section describes the design of \ek: the circuit representation and layered architecture (\cref{sec:architecture}), the modules the library ships (\cref{sec:modules}), the providers, their gradients and the cost of the abstraction (\cref{sec:providers}), and how the library can be extended (\cref{sec:extensibility}). For the full API and implementation details, we refer the reader to the library documentation (\cref{sec:code}).

\subsection{Architecture}
\label{sec:architecture}

\begin{figure*}[t]
    \centering
    \includegraphics[width=\textwidth]{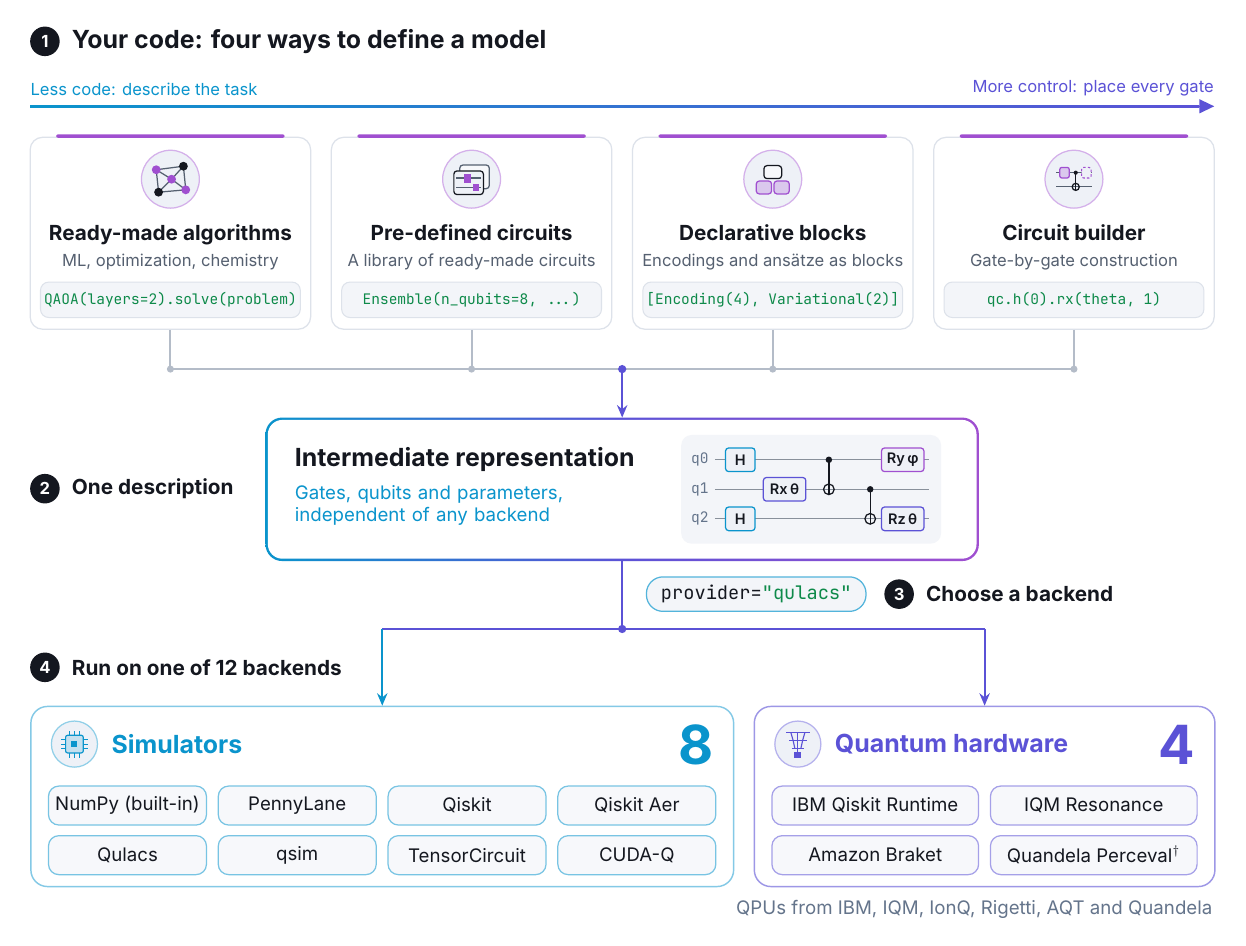}
    \caption{
    Overview of the \texttt{eigenkit} workflow. Models can be defined at different levels of abstraction, compiled into a backend-independent intermediate representation, and executed on one of 12 supported simulation or quantum-hardware backends.
    }
    \label{fig:kit_overview}
\end{figure*}

\ek{} provides a single interface for building quantum circuits and running them on multiple execution backends, both simulators and quantum hardware. To make this possible, the library defines an intermediate representation (IR) of a circuit that decouples construction from execution, as illustrated in Fig.\ref{fig:kit_overview}. The IR describes a circuit as data: the ordered list of gates, the qubits each gate acts on, and the gate parameters, which are either constants or references to trainable weights and data inputs. The IR serves as a circuit schema and stores no numerical values for the weights and inputs: they are supplied at each call. Each provider translates the IR into its own SDK's primitives, following a protocol defined by the library. Users therefore write a circuit once and the library can run it on any supported platform.

Circuits can be constructed in several ways. A user can build a circuit imperatively, defining the qubit set and adding gates sequentially, compose it from predefined building blocks (data encodings, circuit templates and observables) or choose one from a library of predefined ans\"atze. Whatever the entry point, the result is the same IR, so circuits built in different ways behave identically.

The library is organized in three layers: algorithms, circuits, and providers. An \emph{algorithm} represents a complete solution workflow and may combine classical processing, optimization, and one or more quantum circuit evaluations. A \emph{circuit} is the quantum computational unit itself, and a \emph{provider} is responsible for compiling and executing that circuit on a simulator or QPU. Each layer depends only on the interface of the layer below, not on its implementation: an algorithm does not need to know which provider runs its circuits, and a circuit does not know how a provider implements it. Adding a provider, for example, therefore requires no changes to circuits or algorithms.

\subsection{Modules}
\label{sec:modules}

The library is designed to be lightweight and modular, on the principle that a user who works, for example, on optimization algorithms and never trains a hybrid neural network should not have to install large dependencies like PyTorch. The core of the library therefore depends only on NumPy and SciPy. It includes a statevector simulator and a variational optimizer, which is sufficient for most basic tasks. Other frameworks, backends and SDKs are optional: each is installed as an extra and imported only when it is first used, so users install only what their work actually requires. The library currently ships the following modules:
\begin{itemize}
  \item \textbf{core}: This is the circuit layer the rest of the library is built on: the IR, circuit building blocks and observables, and circuit compilation, dispatch, execution and differentiation.
  \item \textbf{qml}: Scikit-learn-compatible quantum estimators: fidelity-based quantum kernels and quantum support-vector classifiers and regressors.
  \item \textbf{torch}: A PyTorch interface that makes circuits differentiable in PyTorch and exposes them as \texttt{nn.Module} layers covering common hybrid patterns (standard angle encoding followed by a variational layer, data re-uploading, chunked input encoding, ensembles of quantum circuits that process different subsets of the input features, parallel quantum--classical branches), together with hybrid quantum--classical neural-network estimators for classification, regression and time-series problems.
  \item \textbf{optimization}: Quadratic Unconstrained Binary Optimization (QUBO) and Ising problem formulations, a Max-Cut constructor, and a Quantum Approximate Optimization Algorithm (QAOA) solver.
  \item \textbf{chemistry}: Fermionic operators, including Jordan--\allowbreak Wigner and Bravyi--\allowbreak Kitaev mappings to qubit Hamiltonians, Hartree--\allowbreak Fock and UCCSD ans\"atze, and a Variational Quantum Eigensolver (VQE) solver. Molecular Hamiltonians are built through \texttt{OpenFermion} and \texttt{PySCF}, with conversion to and from \texttt{OpenFermion} operators.
  \item \textbf{reservoir}: Quantum reservoir computing with several state-carry policies.
  \item \textbf{algorithms}: General-purpose quantum algorithms: the HHL linear-systems solver and an iterative quantum amplitude estimation (IQAE).
  \item \textbf{benchmark}: Tools for comparing quantum and classical estimators on the same dataset across several random seeds, with a set of classical baselines.
  \item \textbf{datasets}: Registry and loaders for standard classification, regression and time-series datasets (scikit-learn-bundled and downloaded), synthetic generators (moons, circles, checkerboard, time series), and dataset summary reports.
\end{itemize}

\subsection{Providers and portability overhead}
\label{sec:providers}

The current version of the library ships providers for a built-in NumPy simulator, PennyLane, Qiskit, Qiskit Aer (Runtime, statevector and matrix-product-state methods), Qulacs, qsim, TensorCircuit, Amazon Braket, CUDA-Q and Perceval for photonics quantum computing. Not every provider offers its own differentiation method. Whenever a provider supplies gradients, for example the adjoint method of the built-in simulator or PennyLane's own differentiation, the library uses them. Otherwise, it falls back to in-house backend-independent parameter-shift rule. As a result, every simulator supports gradient-based training through the same interface, including simulators whose SDKs would otherwise require a hand-written gradient.

Having such an abstraction layer has its cost. A circuit passes through three stages: assembly, compilation and execution. At assembly, all blocks, gates and parameters are collected into a single object. At compilation, the provider builds its native objects, such as a PennyLane \texttt{QNode} or a parametric Qiskit circuit, which are expensive to create. They are built once and reused on every forward and backward pass, so only execution is repeated. To quantify the routing expenses, we measured the per-call and compile-time cost of the library against hand-written code for three providers, that calls the same SDK directly (see \cref{tab:overhead}).

\begin{table*}[t]
  \centering
  \caption{Per-call and compile-time cost of the library relative to hand-written code against the same SDK, on an angle-encoded depth-3 circuit measuring one Pauli operator per qubit. Best of ten measured calls after three discarded warmups with bitwise parity of the results.}
  \label{tab:overhead}
  \begin{tabular}{@{}rlrrrrr@{}}
    \toprule
    $n$ & Backend & Library ($\mu$s) & Native ($\mu$s) & Dispatch ($\mu$s) & Per-call ratio & Compile ratio \\
    \midrule
     4 & PennyLane & 2148.1 & 2122.2 &  25.9 & 1.01$\times$ & 10.32$\times$ \\
     4 & Qiskit    &  715.5 &  682.6 &  32.9 & 1.05$\times$ &  4.73$\times$ \\
     4 & qulacs    &   28.9 &   16.0 &  12.9 & 1.81$\times$ & 17.31$\times$ \\
    \addlinespace
     8 & PennyLane & 3945.7 & 3873.6 &  72.1 & 1.02$\times$ & 11.81$\times$ \\
     8 & Qiskit    & 1389.9 & 1376.4 &  13.5 & 1.01$\times$ &  3.10$\times$ \\
     8 & qulacs    &   68.6 &   52.9 &  15.7 & 1.30$\times$ &  9.79$\times$ \\
    \addlinespace
    12 & PennyLane & 6782.9 & 6660.6 & 122.3 & 1.02$\times$ & 11.25$\times$ \\
    12 & Qiskit    & 3140.9 & 3124.1 &  16.9 & 1.01$\times$ &  2.66$\times$ \\
    12 & qulacs    &  655.3 &  635.3 &  20.0 & 1.03$\times$ &  7.42$\times$ \\
    \bottomrule
  \end{tabular}
\end{table*}

The library adds a few tens of microseconds per call where distinguishable from noise, while the cost of the simulation itself grows with $n$, making it practically negligible. The measurement was performed on 3 different hosts with run-times varying up to 2 times, and dispatch overhead kept the same absolute values, meaning that the cost is a property of dispatch path, not the host. Compile ratios are larger, from 2.5 to 18 times of the original SDK, but in absolute terms the library's compilation takes 1.3--4.3\,ms and is paid once per circuit creation, not per call. This makes library performance almost identical to direct implementation, while providing the benefits of portability and a unified interface.

\subsection{Extensibility}
\label{sec:extensibility}

The library is designed so that most extensions are additive. A new provider is an object that implements the provider protocol: it declares its capabilities (supported gates, qubit limit, shots, batching) and its gradient method, and implements compilation and execution. It can be passed to a circuit directly or registered under a name, and circuits and algorithms need no changes. Each new provider is tested against a NumPy reference to ensure correctness before being incorporated to the library. New building blocks are likewise composed from existing gates without touching the core. Adding a gate is the one change that extends the core: the gate is added to the gate vocabulary and mapped to each provider's SDK primitives. A provider that does not support a gate omits it from its declared capabilities, and circuits that use it are rejected before compilation. Computing models that do not fit the gate-based IR, such as photonic devices, can be added as separate modules alongside the gate-based stack. Such a module reuses the same design, with a provider protocol, a compile/run split and the same estimator interfaces, rather than forcing a new model into the qubit-circuit representation.

%% =====================================================================
\section{Experiments}
\label{sec:experiments}

In this section, we evaluate the practical consequences of provider-neutral execution. We consider two complementary settings. First, we benchmark several simulation backends across different types of quantum workloads. Second, we study execution on real QPUs and show that the most suitable processor can change with the interaction structure of the circuit. Together, these experiments illustrate that backend selection is inherently workload dependent, which motivates the ability to move the same algorithm across execution environments with minimal changes.

\subsection{Simulation experiments}
\label{sec:sim}

In the NISQ era, circuit simulation remains an important execution environment for quantum algorithms. Quantum hardware offers a scale that is inaccessible to classical simulation, but it is limited by noise and by practical throughput. For example, training a hybrid quantum--classical neural network (HQNN) requires thousands of circuit executions, which is usually impractical on hardware because of both cost and queue times. At the qubit counts typically needed for algorithm development, simulators handle this workload on a single machine and provide a fast feedback loop for development and debugging. However, simulators differ in performance depending on the task. For example, the simulator that is fastest for a single circuit execution may be slow for end-to-end HQNN training, where large batches of circuits are evaluated at every step. This raises a natural question: which simulator suits which task? To answer it, we compare the performance of different providers on a range of workloads.

We divide the experiments into two groups: raw circuit execution and application workloads. The circuit families cover basic variational ans\"atze of various widths and depths, executed at several batch sizes. The application workloads are the library's VQE, QAOA and HQNN implementations.

\begin{figure*}[t]
  \centering
  \includegraphics[width=0.94\textwidth]{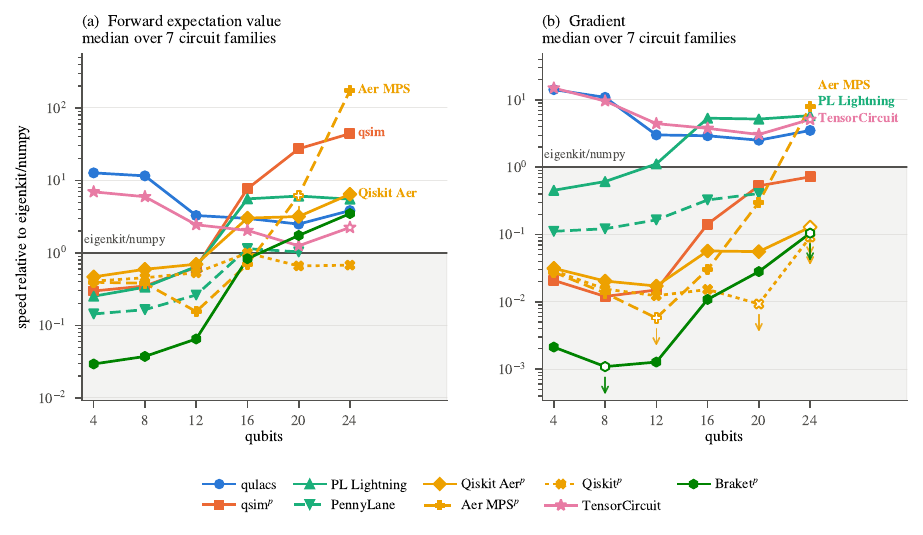}
  \caption{External providers speedup on forward expectation value and gradients computation relative to \ek{} NumPy simulator. The result is median across 7 circuit families. The superscript $p$ marks providers whose gradient comes from the library's parameter-shift fallback rather than a native jacobian. Hollow marker with a downward arrow mean that the median includes at least one timed-out cell, so the plotted value is an upper bound on speed. Hollow markers alone illustrate widths where fewer than 60 percent of families returned a result.}
  \label{fig:speedup}
\end{figure*}

The raw-execution experiments measure execution times for 7 circuit families that vary along three axes: the number of parameters, the amount of entanglement the circuit produces, and how close the circuit unitary to being diagonal. Each point in \cref{fig:speedup} is the median over all circuit families at a given qubit count. The results show that the best choice of simulator is not clear-cut. Simulators that excel at low qubit counts fall behind as the qubit count grows, and vice versa. Tensor-network-based simulators can be either the fastest or the slowest, or fail to run at all, depending on the circuit structure, so choosing one as the default SDK is a gamble. The simulators bundled with hardware SDKs are among the slowest in these measurements. Even for a single circuit execution, selecting the best simulator is not obvious, and the choice becomes even harder for end-to-end workloads.

\begin{figure}[t]
  \centering
  \includegraphics[width=0.9\linewidth]{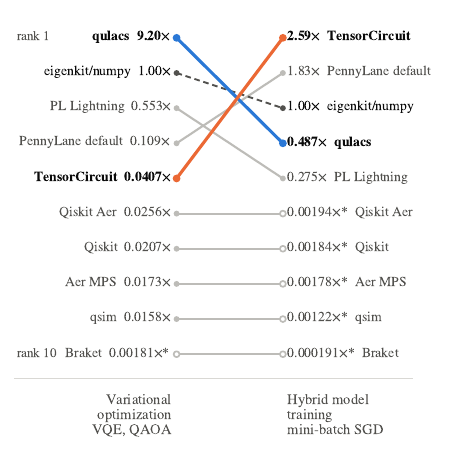}
  \caption{External providers speedup ranking on variational optimization and HQNN training workloads relative to \ek{} NumPy simulator. The asterisk marks a geomean taken over fewer than all configurations due to timeout.}
  \label{fig:ranking}
\end{figure}

The second group of experiments uses algorithms provided by the library: VQE, QAOA and HQNN training. We chose these algorithms because they cover two computational scenarios that differ from the raw circuit execution of the first group. All three are variational optimization algorithms, but VQE and QAOA require far fewer circuit executions per optimization step, while HQNN training evaluates large batches of circuits. This difference shows up in the results (\cref{fig:ranking}): qulacs, the fastest simulator for variational optimization at 9.2$\times$ the NumPy baseline, drops to 0.49$\times$ on HQNN training, a change of almost 20$\times$. The reverse also holds: TensorCircuit, the fastest for HQNN training at 2.6$\times$, runs at only 0.04$\times$ on variational optimization.

Both experiments show that the choice of the best simulator is not obvious and depends strongly on the task. Since much of quantum software development happens in simulation, the choice of SDK can significantly affect development time. Choosing an SDK ties a user to a specific set of simulators, and, as we show, this choice can matter a lot for performance.

\subsection{QPU experiments}
\label{sec:qpu}

\begin{figure*}[t]
  \centering
  \includegraphics[width=0.8\textwidth]{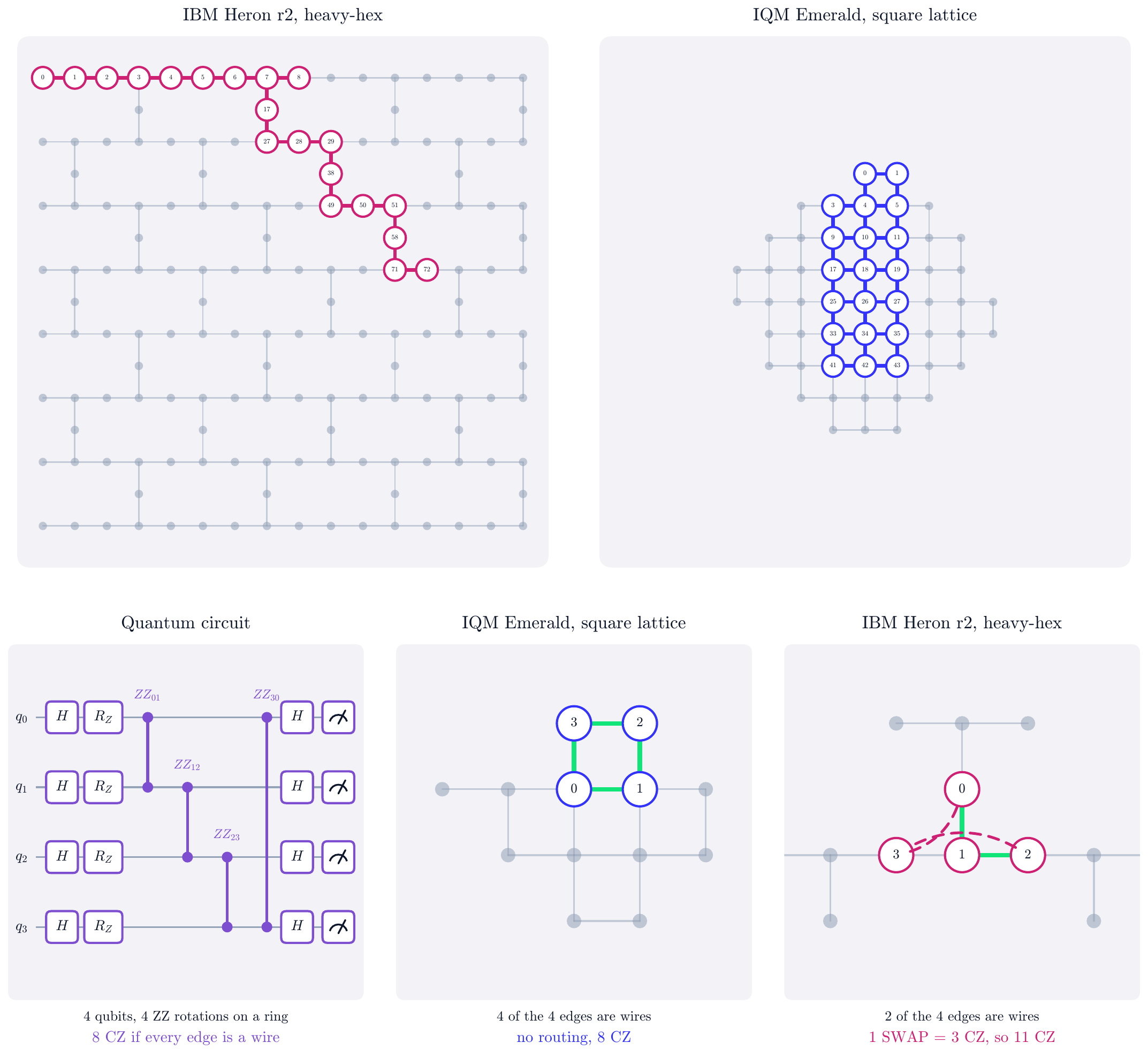}
  \caption{Illustration of topology-dependent routing during transpilation. The top panels show the coupling maps of IBM Heron r2 and IQM Emerald, together with example subgraphs matching the heavy-hex and square-lattice interaction structures considered in this work. The bottom panels illustrate the effect on a four-qubit circuit with $ZZ$ interactions arranged in a ring. On the square-lattice topology, all required interactions can be mapped directly to native couplings, whereas on the heavy-hex topology some logical interactions are not directly available and additional routing is required. In the example shown, one SWAP operation adds three CZ gates, increasing the two-qubit gate count from 8 to 11.}
  \label{fig:routing}
\end{figure*}

The previous experiments show that, even in simulation, the most efficient backend depends on the computational task. On quantum hardware, this choice becomes even more consequential. Given a quantum algorithm, how should one choose the QPU on which to execute it? Different processors offer different trade-offs in qubit count, connectivity, native gates, coherence, and error rates, and there is currently no universally preferable architecture~\cite{darpa_harq}. For example, IBM Heron r2 provides 156 superconducting qubits arranged in a sparse heavy-hexagonal topology, whereas IQM Emerald provides 54 qubits in a more highly connected square lattice~\cite{ibm_heron,iqm_emerald}. The relevant question is therefore not which QPU is best in general, but which one is best matched to the structure of a particular algorithm. As we show below, circuit connectivity can directly determine which processor provides the better execution.

To investigate this effect, we perform a controlled experiment testing whether the most suitable processor can change with the interaction structure of the quantum circuit, even when the circuits belong to the same algorithmic family. We use Quantum Circuit Born Machines (QCBMs)~\cite{liu2018qcbm} constructed from Instantaneous Quantum Polynomial-time (IQP) circuits. A QCBM is a generative model in which a parameterized quantum state $\ket{\psi(\bm{\theta})}$ defines a probability distribution
\begin{equation*}
  p_{\bm{\theta}}(x) = \bigl|\braket{x|\psi(\bm{\theta})}\bigr|^2,
\end{equation*}
from which samples are obtained through measurement. In our case,
\begin{equation}
  \ket{\psi(\bm{\theta})} = H^{\otimes n} D_Z(\bm{\theta}) H^{\otimes n} \ket{0}^{\otimes n},
  \label{eq:iqp}
\end{equation}
where $D_Z(\bm{\theta})$ contains commuting single-qubit $Z$ rotations and two-qubit $ZZ$ interactions applied along the edges of an interaction graph. IQP circuits are particularly interesting for generative modelling because sampling from suitable IQP families is believed to be classically hard under standard complexity-theoretic assumptions~\cite{bremner2011}, although recent work has identified a trade-off between trainability and sampling hardness~\cite{herbst2025}. At the same time, it has been shown that large IQP generative models can be trained classically and subsequently deployed for sampling on quantum hardware~\cite{recio2025,ballo2026}. In this experiment, we consider two different interaction graphs: heavy-hex and a two-dimensional grid. These graphs define genuinely different circuit families and can therefore represent different sets of probability distributions even in the ideal noiseless setting. For the construction considered here, the grid-2D ansatz has higher expressivity~\cite{domingo2026}, potentially allowing it to represent more complex distributions, although increased expressivity can also reduce trainability and lead to flatter optimization landscapes~\cite{holmes2022}. Different IQP architectures can therefore be preferable for different tasks. Crucially, the two interaction graphs also align naturally with the coupling maps of the hardware considered here: the heavy-hex circuit matches the topology of IBM's Heron processor, whereas the grid-2D circuit matches the square-lattice connectivity of IQM Emerald.

When the circuit interaction graph does not match the hardware coupling map, compilation requires additional routing, typically through SWAP operations or equivalent sequences of native two-qubit gates. This effect is illustrated in \cref{fig:routing} for a simple four-qubit ring. The circuit can be embedded directly in the square-lattice connectivity of IQM Emerald, whereas the same interaction pattern is not directly supported by the heavy-hex connectivity of IBM Heron and requires additional routing. In the example shown, a single SWAP increases the number of CZ gates from 8 to 11. Since two-qubit operations remain a major source of error on current QPUs, this additional routing overhead can directly affect execution quality. Our experiment therefore evaluates how the compatibility between circuit and hardware topology influences the quality of the resulting quantum computation.

We evaluate heavy-hex and grid-2D circuits at $n=20$ and $n=54$ qubits on both processors, giving 8 hardware executions in total. Each circuit is executed using the same parameter realization on both devices with \num{8000} shots, and the interaction graphs are extracted directly from the physical coupling maps of the corresponding processors. To quantify execution quality, we compare the experimentally estimated Pauli-$Z$ expectation values,
\begin{equation}
  \langle Z_\omega \rangle_{\mathrm{QPU}} = \frac{1}{S} \sum_{s=1}^{S} (-1)^{\sum_{i\in\omega} b_i^{(s)}},
  \label{eq:zexp}
\end{equation}
with their ideal noiseless values $\langle Z_\omega\rangle_{\mathrm{ideal}}$, where $\omega$ denotes a subset of qubits and $b^{(s)}$ is the measured bit string for shot $s$. Importantly, these ideal values do not require simulation of the complete $2^n$-dimensional state vector. The commuting structure of the IQP circuit allows low-weight $Z$ expectation values to be computed exactly and classically from the circuit interaction graph and its parameters~\cite{recio2025}. For an observable supported on $\omega$, this calculation scales as $\mathcal{O}(2^{|\omega|}\,|\partial\omega|)$, where $\partial\omega$ denotes the neighbours of $\omega$ in the interaction graph. We restrict the analysis to observables of weight at most four and retain the 64 expectation values with the largest $|\langle Z_\omega\rangle_{\mathrm{ideal}}|$, allowing exact reference values to be obtained efficiently even for the 54-qubit circuits.

\begin{table}[t]
  \centering
  \caption{Experimental results on IBM Heron and IQM Emerald. Lower RMSE and higher slope indicates better agreement with the ideal expectation values.}
  \label{tab:qpu}
  \begin{tabular}{@{}rllrr@{}}
    \toprule
    $n$ & Ansatz & Device & $\alpha$ & RMSE \\
    \midrule
    20 & heavy-hex & \textbf{IBM} & \textbf{0.9192} & \textbf{0.0339} \\
    20 & heavy-hex & IQM          & 0.9005          & 0.0402 \\
    20 & grid-2D   & IBM          & 0.7952          & 0.0457 \\
    20 & grid-2D   & \textbf{IQM} & \textbf{0.8509} & \textbf{0.0329} \\
    \addlinespace
    54 & heavy-hex & \textbf{IBM} & \textbf{0.7923} & 0.1716 \\
    54 & heavy-hex & IQM          & 0.7427          & \textbf{0.1355} \\
    54 & grid-2D   & IBM          & 0.7646          & 0.0955 \\
    54 & grid-2D   & \textbf{IQM} & \textbf{0.8482} & \textbf{0.0660} \\
    \bottomrule
  \end{tabular}
\end{table}

We characterize the agreement between hardware and ideal execution using two complementary metrics. First, we compute the least-squares slope
\begin{equation}
  \alpha = \frac{\sum_\omega \langle Z_\omega\rangle_{\mathrm{QPU}} \cdot \langle Z_\omega\rangle_{\mathrm{ideal}}}{\sum_\omega \langle Z_\omega\rangle_{\mathrm{ideal}}^2}.
  \label{eq:slope}
\end{equation}
If hardware noise approximately attenuates the ideal expectation values as $\langle Z_\omega\rangle_{\mathrm{QPU}} \simeq \alpha \langle Z_\omega\rangle_{\mathrm{ideal}}$, then $\alpha$ quantifies how much of the ideal signal magnitude is preserved, with $\alpha=1$ corresponding to the ideal case. Second, we report the root mean squared error (RMSE), which directly measures the deviation between the measured and ideal expectation values, with lower values indicating better agreement. The slope therefore captures the preservation of the ideal signal scale, while the RMSE captures the absolute execution error.

The results in \cref{tab:qpu} show a clear dependence on the compatibility between the circuit interaction graph and the hardware coupling map. We first consider the 20-qubit heavy-hex circuit. On IBM, this circuit is natively compatible with the heavy-hex topology, while on IQM its relatively low connectivity allows the transpiler to find a suitable embedding within the larger 54-qubit square lattice. Consequently, both processors execute the circuit with the same 38 two-qubit gates and no routing overhead. Their performance is correspondingly similar; with routing effects effectively removed, the remaining difference reflects device-level noise characteristics. The situation changes for the grid-2D circuits. Their degree-4 interaction structure maps naturally onto IQM's square lattice, but cannot be embedded directly into IBM's degree-3 heavy-hex topology, making additional routing unavoidable. At 20 qubits, this increases the two-qubit gate count from the native 60 gates on IQM to 107 on IBM, while at 54 qubits the difference grows from 170 to 427 gates. In both cases, IQM achieves better agreement with the ideal execution, consistent with the substantially lower routing overhead. Finally, the 54-qubit heavy-hex circuit reverses the situation. IBM executes the circuit natively using 118 two-qubit gates. Although IQM has sufficient local connectivity to embed smaller heavy-hex circuits, at $n=54$ all physical qubits are occupied, leaving no freedom to select a more suitable subset of the processor. The full IQM coupling map does not contain the required heavy-hex interaction graph natively, so routing increases the two-qubit gate count to 206. Accordingly, IBM achieves the larger slope. Overall, these results show that the preferred processor changes with the circuit topology: additional routing induced by a mismatch between logical and physical connectivity can substantially degrade execution quality.

These results illustrate why provider-neutral execution is useful in practice. The most suitable QPU depends not only on the hardware itself, but also on how well the circuit maps onto that hardware. With \ek, the same algorithm can be executed across different simulators and QPUs by changing only the provider configuration, as illustrated in \cref{lst:iqp}. This makes it easier to compare backends and select the one that is better suited to a particular workload without modifying the high-level implementation. Together with the simulation results above, this shows that backend choice is inherently workload dependent.
\clearpage
\begin{eopanel}[left=3mm,right=3mm,unbreakable]
\begin{lstlisting}[caption={Example of a provider-neutral execution of the same IQP circuit across simulators and quantum hardware backends.},label={lst:iqp},captionpos=b,aboveskip=0pt,belowskip=0pt]
from eigenkit import (Circuit, CircuitBuilder, IQMConfig,
    Observable, QiskitRuntimeConfig)

# Define IQP circuit: H^n, a diagonal ZZ layer, H^n
def iqp(n, edges, theta):
    builder = CircuitBuilder(n)
    for q in range(n):
        builder.h(q)
    for (i, j), angle in zip(edges, theta):
        builder.pauli_exponential((("Z", i), ("Z", j)), angle)
    for q in range(n):
        builder.h(q)
    return builder.build()

# Build circuit and observables
spec = iqp(4, [(0, 1), (1, 2), (2, 3), (3, 0)],
           [0.7, 1.1, 0.4, 1.9])
observables = [Observable.from_string(s)
               for s in ("Z0 Z1", "Z1 Z2", "Z0 Z1 Z2 Z3")]

# Define backends
providers = {
    "exact": "numpy",
    "qulacs": "qulacs",
    "IBM": QiskitRuntimeConfig(backend_name="ibm_kingston",
        shots=8192, primitive="sampler", seed_transpiler=11),
    "IQM": IQMConfig(quantum_computer="emerald", shots=8192,
        use_metrics=True, seed_transpiler=11)}

for name, provider in providers.items():
    circuit = Circuit(spec, observables=observables,
                      provider=provider)
    print(name, circuit(np.zeros(0)))
\end{lstlisting}
\end{eopanel}

%% =====================================================================
\section{Limitations and future work}
\label{sec:limitations}

Like any provider-neutral framework that integrates multiple SDKs, \ek{} necessarily operates within the capabilities of the underlying SDKs and hardware platforms. Therefore, not every provider accepts every circuit: some providers do not support all gates or parameter expressions. For example, the CUDA-Q provider does not accept sums or products of parameters in gate angles. The library detects such incompatibilities and rejects unsupported circuits with an explicit error rather than executing them incorrectly, in which case the user may choose a different provider. Optional features also differ between providers. Shot-based sampling, for instance, is available on PennyLane, Braket, CUDA-Q and the hardware providers, some of the other simulators currently return exact expectation values only. We aim to keep the execution interface uniform wherever possible, while exposing provider-specific capability differences explicitly.

The empirical evaluation presented in this paper also has a defined scope. All simulator timings were obtained on a single x86 machine, pinned to one thread, and GPU or multi-threaded execution may lead to different relative performance across providers. The QPU comparison uses a single parameter realization per circuit and one execution per device, and therefore does not capture run-to-run variation or calibration drift; the differences in \cref{tab:qpu} should consequently be interpreted as evidence of the effect of topology and routing in the performance of a particular circuit rather than as a statistically resolved ranking of the devices.

Future work includes expanding the set of supported providers, modules, algorithms and blocks, as well as exploring new directions such as noise modelling and analogue quantum computing. Another priority is usability: extending support for common programming patterns and utility functions to streamline application development. We also plan closer integration with established frameworks such as Qiskit and PennyLane, not only at the execution layer but also for circuit construction, for example through circuit import and export. The library's development is currently driven by application needs, and its modular architecture is intended to support continued extension as new use cases and execution environments arise.

%% =====================================================================
\section{Conclusions}
\label{sec:conclusions}

We presented \ek, a provider-neutral quantum computing library designed to separate the definition of a quantum algorithm from the backend used to execute it. The library provides a common interface across heterogeneous simulators and quantum hardware, together with backend-independent gradient calculation, and integration with standard scientific-computing and machine-learning workflows. Its goal is not to replace existing quantum software frameworks, but to provide an abstraction layer that makes quantum algorithms easier to move, compare, and integrate across them.

The experiments show why this flexibility is relevant in practice. Across simulators, performance changes substantially with circuit size, structure, batching, and application workload, so no single backend is consistently preferable. On quantum hardware, which processor is better suited to a given algorithm depends not only on hardware constraints such as system size and error rates, but also on how well the topology of the quantum algorithm matches the physical connectivity of the processor.

Together, these results motivate a workflow in which backend choice can remain flexible throughout algorithm development and deployment. By allowing the same high-level implementation to be evaluated across different execution environments with minimal changes, \ek{} aims to make heterogeneous quantum computing more accessible to systematic experimentation, benchmarking, and application development.

%% =====================================================================
\section{Code availability}
\label{sec:code}

The library is \ek, released under the Apache License 2.0 at \url{https://github.com/eigenorbit/eigenkit}. The version measured here is 1.0.0, tagged \texttt{v1.0.0}; \texttt{pip install eigenkit} installs the numpy-only core, and the PyTorch, scikit-learn and per-backend integrations are opt-in extras.

\bibliography{refs}

\end{document}